# A model for the optical flares from the Galactic transient SWIFT J195509+261406

XU Ming & HUANG Yong-Feng[†]

Department of Astronomy, Nanjing University, Nanjing 210093, China

**The Galactic hard X-ray transient SWIFT J195509+261406 was first observed as a gamma-ray burst GRB 070610. Within 3 days after the burst, more than forty optical flares had been observed. Here we propose that this peculiar event should be associated with a white dwarf. The hard X-ray burst itself may be triggered by the collision between two planets orbiting the white dwarf. Some cracked fragments produced in the collision then fall onto the surface of the white dwarf in several days, giving birth to the observed optical flares via cyclotron radiation. Our model can satisfactorily explain the basic features of the observations.**

gamma-ray bursts, white dwarfs, cyclotron radiation

At 20:52:26 UT on June 10 2007, the Burst Alert Telescope onboard the *Swift* satellite detected GRB 070610 / SWIFT J195509+261406. The prompt emission has a duration of $T_{90} = 4.6 \pm 0.4 s$. The spectrum in 15-150keV range is best fit by a power-law function with a photon index of $\Gamma = 1.76 \pm 0.25$, which gives a flux of $F = (2.4 \pm 0.4) \times 10^{-7} ergs \cdot cm^{-2}$ [1]. X-ray afterglows were also observed, but with curious rapid variabilities and flares[2]. In the following three days, more than 40 intense optical flares were observed from this source, with typical durations of ~100 s [2-4]. Thus the overall behavior of SWIFT J195509+261406 is quite different from that of a classical gamma-ray burst (GRB). Since SWIFT J195509+261406 locates in the Galactic plane, Kann et al. argued that it should be a Galactic transient, but not a classical GRB [5]. In fact, SWIFT J195509+261406 is now classified as a hard X-ray transient.

In view of these unusual properties of SWIFT J195509+261406, we here propose a new model for it. We suggest that the hard X-ray burst should be triggered by the collision between two planets orbiting a white dwarf (WD). The cracked fragments produced in the collision then fall one by one onto the surface of the WD in several days, giving birth to the observed optical flares.

## 1  Hard X-ray burst from two-planet collision

The observations of planet system such as PSR B1257+12 imply that planets could survive a supernova and remain bound to the new-born compact star[6]. Such planets may be left in eccentric and coplanar orbits. It is possible that the orbits will intersect and these survived planets will collide at a late time[7]. Here we consider a system of two planets surrounding a WD. The mass of the WD is taken as $M = 1 M_\odot$, where $M_\odot$ is the solar mass. We assume that the two planets collide at the position with a distance of $R_0 = 10^{12} cm$ away from the WD. For simplicity, we further assume that the two planets are equal in mass with $m_p = 10^{27} g$. Then the total kinetic energy of the collision will approximately

Received Apr 30, 2009; accepted Jun 18, 2009
doi:
[†]Corresponding author (email: hyf@nju.edu.cn)
Supported by the National Natural Science Foundation of China (Grant No. 10625313) and the National Basic Research Program of China (973 Program, grant 2009CB824800) be the kinetic energy of the two planets surrounding the WD, i.e. $E_{coll} \sim GMm_p / R_0 \sim 1.3 \times 10^{41} ergs$.

Note that the observed isotropic energy of the hard X-ray burst of SWIFT J195509+261406 in 15-150keV band is

$$E_{b,iso} = 4\pi d_L^2 F \sim 2.9 \times 10^{39} ergs \left(\frac{d_L}{10kpc}\right)^2 \left(\frac{F}{2.4 \times 10^{-7} erg \cdot cm^{-2}}\right), \quad (1)$$

where $d_L$ is the distance between the source and us. Since $E_{coll} \gg E_{b,iso}$, we see that the energy released by the collision is large enough to account for the observed hard X-ray burst.

## 2 Falling of fragments onto WD surface

The violent collision between two planets will produce many small fragments with different sizes and velocities. Some fragments will be captured by the WD and finally fall onto the compact star. Let us consider a typical fragment with a mass of $m = 10^{22} g$. For simplicity, we assume that it is a cylinder with a radius of $r = 8 \times 10^6 cm$ and a length of $2r = 1.6 \times 10^7 cm$. We take the magnetic field and the radius of the WD as $B = 1.3 \times 10^8 G$, $R = 5 \times 10^8 cm$.

The falling of a fragment onto WD surface must satisfy the conservation of both energy and angular momentum. Then we can obtain the radial velocity of a falling fragment as

$$v_r(x) = \sqrt{v_{r0}^2 + v_{\theta 0}^2 + 2GM\left(\frac{1}{x} - \frac{1}{R_0}\right) - v_{\theta 0}^2 \frac{R_0^2}{x^2}}, \quad (2)$$

where $v_{r0}$ and $v_{\theta 0}$ are the initial radial and tangential velocity at the collision position ($R_0 = 10^{12} cm$) respectively, $x$ is the distance between the WD and the fragment. As long as the initial velocity of the falling fragment satisfies the condition of $v_0^2 = v_{r0}^2 + v_{\theta 0}^2 < 2GM / R_0$, the fragment will fall onto the WD surface.

The timescale for a fragment to fall onto the WD surface from the collision position can be calculated by

$$t = \int_{R_0}^{R} \frac{dx}{v_r(x)}, \quad (3)$$

Assuming the initial velocity as $v_0^2 = 1.5 \times 2GM / R_0$, $v_{r0}^2 = 0.2v_0^2$ and $v_{\theta 0}^2 = 0.8v_0^2$, we can obtain that the falling timescale is $3 \times 10^4 s$.

For the violent collision, the velocity of the fragments may have a wide range, so the fragments will fall onto the WD surface in several days. We see that our model can give a natural explanation for the time range of the observed optical flares.

The impact of a solid object onto a neutron star (NS) has been studied by Colgate et al.[8]. Near the NS surface, the volume of the object will be compressed by the strong gravitational field. In the case of a WD, the tidal breakup radius is $R_b = (\rho r^2 MG / S)^{1/3} \sim 1.4 \times 10^{10} cm$, where $\rho$ is the density of the fragment and $S$ is its shear strength. So the elongate factor of the fragment in radial direction is $f_r = \frac{1}{5}[4(R/R_b)^{-1/2} + (R/R_b)^2] \sim 4.2$ and the compression factor in tangential direction is $f_\theta = 2(R/R_b)^{1/2} - (R/R_b) \sim 0.34$ when the fragment approaches the WD surface. We can neglect the effect of the WD magnetic field in the falling phase, since the energy density of the fragment far exceeds the magnetic energy density.

## 3 Optical flares from the impacts

The observed optical flares are mostly in I-band, with maximal flux of $F_v \sim 19^m$ and duration of $\Delta t \sim 100s$. Since the source locates in the Galactic plane, the reddening could be large as $A_I \sim 6^{m\,[4,5]}$. So a typical flare has an intrinsic flux of $F_v \sim 13^m$ ($\sim 14.2 mJy$). The maximal luminosity of the flares can be estimated by

$$L = 4\pi d_L^2 \int F_v d\nu \sim 2.1 \times 10^{36} ergs \cdot s^{-1} \left(\frac{d_L}{10kpc}\right)^2 \left(\frac{F_v}{14.2mJy}\right)\left(\frac{\Delta \nu}{10^{14} Hz}\right), \quad (4)$$

where $\Delta \nu$ is the width of band. So the released energy of a typical flare is $\Delta E_f \sim \frac{1}{2} L \Delta t \sim 1.1 \times 10^{38} ergs$. Theoretically, the gravitational energy release from the impact of a fragment onto the WD surface is

$$E_g = \frac{GMm}{R} \sim 2.7\times 10^{39}\, ergs \\ \left(\frac{M}{M_\odot}\right)\left(\frac{m}{10^{22}\,g}\right)\left(\frac{R}{5\times 10^8\,cm}\right)^{-1}. \quad (5)$$

Again we see that this energy will be large enough to explain the observed flares. Below, we investigate the impact process in more detail.

### 3.1 Impact depth

From the hydrostatic equilibrium equation ($dp/dr = -\rho_s GM/R^2$) and the equation of state ($p = K\rho_s^\Gamma$), we derive the density structure in the surface layer of the WD as $\rho_s = \left(\frac{2GMd}{5KR^2}\right)^{3/2}$, where $d$ is the depth below the WD surface, $K = 3\times 10^{12}$ and $\Gamma = 5/3$ [9]. The fragment will stop when its swept-up mass in the WD equals its own mass. So the impact depth is

$$d = \frac{5}{2}\left(\frac{m^2 R^6 K^3}{\pi^2 f_\theta^4 r^4 G^3 M^3}\right)^{1/5} \sim 1.3\times 10^6\, cm. \quad (6)$$

At this depth, the pressure is $P_{deg} = K\rho_s^\Gamma \sim 2.3\times 10^{17}\, ergs\cdot cm^{-3}$. It is much less than the impact energy density $P_{in} = E_g/\pi f_\theta^2 r^2 d \sim 2.3\times 10^{19}\, ergs\cdot cm^{-3}$. So the fragment will explode and produce a fireball. The expansion velocity of the fireball is

$$v_{exp} \sim \sqrt{\frac{\gamma P_{in}}{\bar\rho}} \sim 6.7\times 10^8\, cm\cdot s^{-1}, \quad (7)$$

where $\bar\rho$ is the average density of swept-up matter and $\gamma \sim \frac{5}{3}$ is the adiabatic index.

### 3.2 Cyclotron radiation of fireball electrons

The expanding fireball contains a lot of electrons. They rotate in the WD magnetic field and emit photons through cyclotron radiation. In the gravitation field of the WD, the expanding fireball will decelerate and fall back. The ascending height of the fireball is

$$h = \left(\frac{1}{R} - \frac{v_{exp}^2}{2GM}\right)^{-1} \sim 3.2\times 10^9\, cm. \quad (8)$$

The ascending timescale is $\sim 17.2\,s$ (Eq. (3)). The duration of the flare is determined by the maximum of the cooling and ascending timescales, which is consistent with the observed values.

In the WD magnetic field, the cyclotron frequency of electrons is

$$\nu_L = \frac{eB}{2\pi m_e c} \sim 3.6\times 10^{14}\, Hz \left(\frac{B}{1.3\times 10^8\, G}\right). \quad (9)$$

It is in optical range, in agreement with observations. The average radiation power of a single electron is

$$\bar P = \frac{4}{9c} r_0^2 v^2 B^2 \sim 2.4\times 10^{-2}\, ergs\cdot s^{-1} \\ \left(\frac{v}{6.7\times 10^8\, cm/s}\right)^2 \left(\frac{B}{1.3\times 10^8\, G}\right)^2, \quad (10)$$

where $r_0$ is the classical radius of electrons, $v$ is electron velocity. In order to produce the observed luminosity of $L \sim 2.1\times 10^{36}\, ergs\cdot s^{-1}$, the electron number should be $L/\bar P \sim 8.8\times 10^{37}$. It is far less than the real number of electrons included in the fireball, i.e., $\sim 10^{22}\,g/1.67\times 10^{-24}\,g \sim 6.0\times 10^{45}$. In fact, the radiation timescale of a typical electron is only $\sim \frac{1}{2}m_e v_{exp}^2/\bar P \sim 1.7\times 10^{-8}\,s$, which means it will only travel a distance of ~11 cm before losing most of its energy. So, the duration of the optical flare should be determined by the ascending timescale of the ejected material. We can imagine that in tens of seconds, the ejected material will continuously fall back onto the WD, being heated and producing the observed flare. So, at any moment of the flare, only a small portion of electrons are heated and emitting photons via cyclotron radiation. In short, we see that the observed luminosity of the flares can be explained in our scenario.

## 4 Conclusions and discussion

We have presented a model for the amazing galactic transient SWIFT J195509+261406. In our scenario, the hard X-ray burst is resulted from the collision between two planets around a WD. The collision also gives birth to many fragments with different sizes and initial ve-locities. Some fragments fall onto the WD in a few days. The impacts between these fragments and the WD will produce energetic fireballs. High velocity electrons in the fireball move in the WD magnetic field. They emit optical photons through cyclotron radiation and produce the observed optical flares.

We have shown that our model can explain many of the observed features, such as the energetics of the hard X-ray burst and the time range of the optical flares. The predicted durations, energy releases and characteristic photon frequency of the flares are also consistent with observations. In our model, most flares may happen at ~1.1 days after the burst trigger, which correspond to the falling of those fragments with nearly zero radial velocity onto the WD surface.